\documentclass[letterpaper,onecolumn,11pt,unpublished]{quantumarticle} %
\pdfoutput=1
\usepackage[english]{babel}
\usepackage[T1]{fontenc}

\usepackage{amsfonts,amsmath,amssymb,amsthm,graphicx,float,hyperref,comment,color,bm}

\usepackage{algcompatible}
\usepackage[ruled,vlined]{algorithm2e}
\usepackage[title]{appendix}
\usepackage{tikzscale}

\usepackage{framed}

\usepackage{authblk}
\usepackage{siunitx}

\usepackage{enumitem}

\newcommand{\rgl}{\rangle}

\newcommand{\psib}{\boldsymbol{\psi}}

\newcommand{\phib}{\boldsymbol{\phi}}

\usepackage{sansmath}

\usepackage{todonotes}

\usepackage{soul}

\title{Deterministic and Bayesian Characterization of Quantum Computing Devices}
\author{Zhichao Peng}
\email{pengzhic@msu.edu}
\affiliation{Department of Mathematics, Michigan State University}
\author{Daniel Appel\"o}
\email{appelo@vt.edu}
\affiliation{Department of Mathematics, Virginia Tech}
\author{N.~Anders Petersson}
\email{ petersson1@llnl.gov}
\affiliation{Center for Applied Scientific Computing, Lawrence Livermore National Laboratory}
\author{Mohammad Motamed}
\email{motamed@unm.edu}
\affiliation{Department of Mathematics and Statistics, University of New Mexico}
\author{Fortino Garcia}
\email{ fortino.garcia@cims.nyu.edu}
\affiliation{Courant Institute of Mathematical Sciences, New York University}
\author{Yujin Cho}
\email{cho25@llnl.gov}
\affiliation{Quantum Coherent Device Physics, Lawrence Livermore National Laboratory}

\begin{document}

\maketitle

\begin{abstract}
Motivated by the noisy and fluctuating behavior of current quantum computing devices, this paper presents a data-driven characterization approach for estimating transition frequencies and decay times in a Lindbladian dynamical model of a superconducting quantum device. The data includes parity events in the transition frequency between the first and second excited states. A simple but effective mathematical model, based upon averaging solutions of two Lindbladian models, is demonstrated to accurately capture the experimental observations. A deterministic point estimate of the device parameters is first performed to minimize the misfit between data and Lindbladian simulations. These estimates are used to make an informed choice of prior distributions for the subsequent Bayesian inference. An additive Gaussian noise model is developed for the likelihood function, which includes two hyper-parameters to capture the noise structure of the data. The outcome of the Bayesian inference are posterior probability distributions of the transition frequencies, which for example can be utilized to design risk neutral optimal control pulses. The applicability of our approach is demonstrated on experimental data from the Quantum Device and Integration Testbed (QuDIT) at Lawrence Livermore National Laboratory, using a tantalum-based superconducting transmon device.


\end{abstract}



\section{Introduction}

Quantum optimal control (QOC) provides an attractive approach for finding the best possible control pulses to realize efficient operations on quantum devices~\cite{glaser2015training,koch2022quantum}. 
However, optimal control is only meaningful if the parameters in the dynamical model of the device are accurately characterized. In the current Noisy Intermediate-Scale Quantum (NISQ) \cite{bharti2022noisy} era, where device parameters are noisy and fluctuate in time, QOC must therefore be combined with quantum characterization to provide reliable outcomes. 
The conventional experimental approach for determining device parameters, e.g., transition frequencies, $T_1$ and $T_2$ decoherence times, involves curve fitting techniques and/or frequency domain analysis \cite{qiskit_url_calibration,peterer2015coherence,krantz2019quantum}. In addition, there are characterization approaches that directly work with the Lindbladian~\cite{lindblad1976generators} or Hamiltonian matrices in the governing equations, see for example the works by Wiebe, Granade and co-authors~\cite{PhysRevA.89.042314,Wiebe_2015,PhysRevLett.112.190501}. In their approach the characterization algorithms can scale to large systems, assuming that an already large and trustworthy (characterized) device is available. A good survey of current Hamiltonian and Lindbladian learning techniques is presented in the recent paper by Wittler et al.~\cite{wittler2021integrated}. In a related study, an optimal control based pulse-level compiler with hardware calibration was developed by Cho et al.~\cite{cho2023direct}. We also note that several software packages have been developed for characterization and control of quantum devices~\cite{wittler2021integrated,roy2022software,teske2022qopt,ball2021software}.

\begin{figure}[thb]
\centering
  \includegraphics[width=0.4\textwidth]{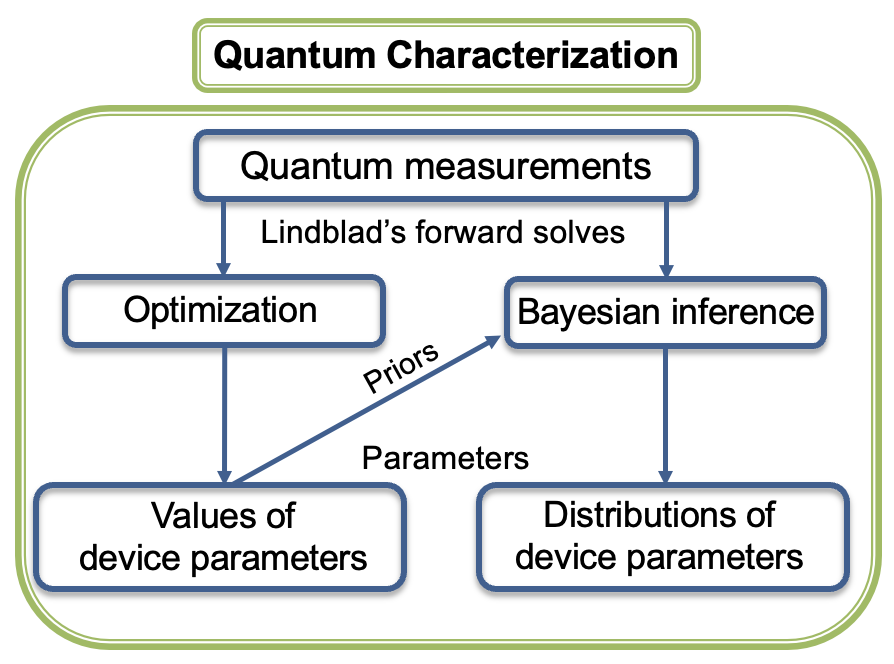}
\caption{Workflow of the deterministic and Bayesian characterizations.}\label{fig:GLOQ}
\end{figure}

In this work, we present both a deterministic and a Bayesian characterization approach based on Lindblad's master equation. 
The characterization procedure, outlined in Figure \ref{fig:GLOQ}, begins by collecting experimental data. 
In the deterministic setting, the parameter inference problem is stated as a Lindblad-constrained optimization problem in which the 
the misfit between the data and solutions of Lindblad's master equation is minimized to find optimal values of the system parameters. 
%
In the Bayesian setting, we infer the probability distribution of system parameters using Bayes' rule and a Markov Chain Monte-Carlo (MCMC) sampling algorithm; Section \ref{sec:bayesian}.
It is to be noted that while the deterministic approach finds point values of device parameters, the Bayesian framework infers their probability distributions. 
Importantly, these distributions account for the noise in quantum measurements and can hence be used for more accurate simulations of quantum systems, as well as enhanced understanding of the underlying physical processes. The distributions can also be useful for designing risk neutral optimal control pulses~\cite{ge2021risk,petersson2021optimal}, for example using the {\tt Juqbox.jl}~\cite{Juqbox-software} software, for improved resilience to noise in the device parameters. The applicability of our approach is demonstrated on experimental results from the Quantum Device and Integration Testbed (QuDIT) at Lawrence Livermore National Laboratory, using a tantalum-based superconducting transmon device~\cite{place2021new}. 
The experimental observations demonstrate a complex noise structure that we attribute to parity events~\cite{riste2013millisecond,martinez2023noise}. This phenomena is observed in the transition frequency between the first and second excited states during a Ramsey experiment. To capture parity events in the simulations we use a simple but effective mathematical model, based upon averaging solutions of two Lindblad equations. This model turns out to be crucial for fitting data from the Ramsey experiment; also see~\cite{peterer2015coherence}. 

The remainder of the paper is organized as follows. Section \ref{sec:prelim} describes the quantum computing hardware, the experimental protocols, and important features of the data. Section \ref{sec:model} presents the governing equations, numerical methods, and the modeling of parity events. The deterministic and Bayesian characterization techniques are introduced in Section~\ref{sec_least-squares} and Section~\ref{sec:bayesian}, respectively, where also the modeling results are compared to experimental data. 
Conclusions are given in Section \ref{sec:conclusion}.

\section{The quantum hardware and experimental protocols}\label{sec:prelim}

In order for quantum simulations to realistically model the hardware and for optimal control to be successful, it is crucial to accurately characterize the device parameters, including transition frequencies and $T_1$ and $T_2$ decoherence times. These parameters determine the coefficients in Lindblad's equation \eqref{eq:lindblad} that, in turn, govern the dynamical model upon which the optimal control relies. 
To accomplish accurate characterization, we need to first collect experimental data from Ramsey and energy decay experiments, followed by inverting for, or inferring probability distributions of, the device parameters. 
%
%
%
In this section, we briefly review the quantum hardware used in the present study and the experimental protocols for the Ramsey and energy decay experiments, followed by a discussion of the experimental results. 

\subsection{Quantum hardware\label{sec:hardware}}

The experiments in this study are performed on a tantalum-based superconducting transmon device~\cite{place2021new}, within the Quantum Device and Integration Testbed (QuDIT) at LLNL. In principle, a transmon can support many energy levels. 
However, on this system only the three lowest levels, corresponding to the states $|0\rgl$, $|1\rgl$ and $|2\rgl$, can be reliably measured. In the following, the transmon device will be referred to as the "qudit". 

To control the qudit and read out results, an IQ mixer is used to generate microwave pulses. The IQ mixer takes an intermediate-frequency (IF) envelope signal and mixes it with a local oscillator (LO) base signal. The IF signal consists of in-phase (I) and quadrature (Q) components, with a frequency content of 50 - 150 MHz. The frequency of the LO base signal is fixed and typically around a few GHz.  For read-out of the quantum state, a measurement signal in the GHz range is down-converted to a signal at $\approx 50$ MHz, via an IQ mixer. This demodulated signal is further analyzed with an OPX instrument (from Quantum Machines) to distinguish between the different quantum states. 

\subsection{Experimental protocols\label{sec:exps}}

Because only the first three states ($|0\rgl$, $|1\rgl$ and $|2\rgl$) can be accurately measured on the qudit, we limit the experiments to determine the $0 \leftrightarrow 1$ and $1 \leftrightarrow 2$ transition frequencies as well as the corresponding decay and dephasing times.

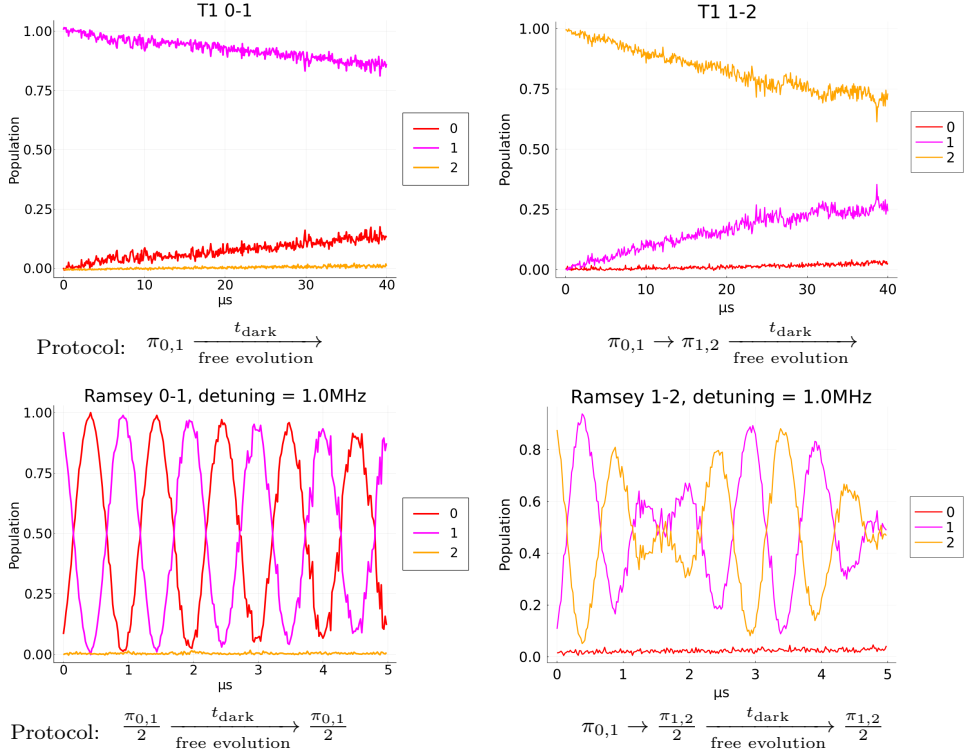
\begin{figure}[thb]
  \begin{center} 
  \begin{tikzpicture}
  \node (ImgT1){\includegraphics[width=0.4\textwidth]{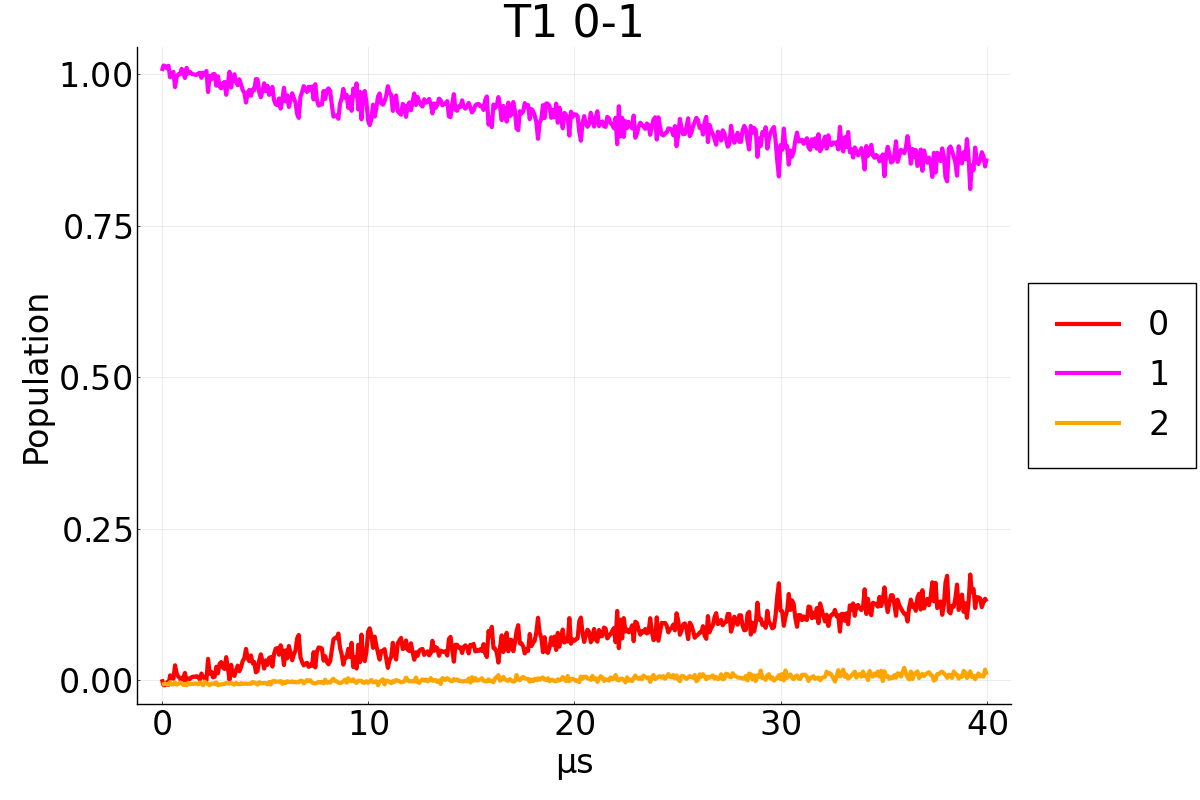}};
  \node(T1)[below=of ImgT1,yshift=+1.2cm]{\scriptsize$\pi_{0,1}\xrightarrow[\textrm{free evolution}]{t_{\textrm{dark}}}$};
 \node(T1)[left=of T1,yshift=-0.05cm,xshift=1.0cm]{\scriptsize Protocol:};
  \end{tikzpicture}
    \begin{tikzpicture}
  \node (ImgT1-12){\includegraphics[width=0.4\textwidth]{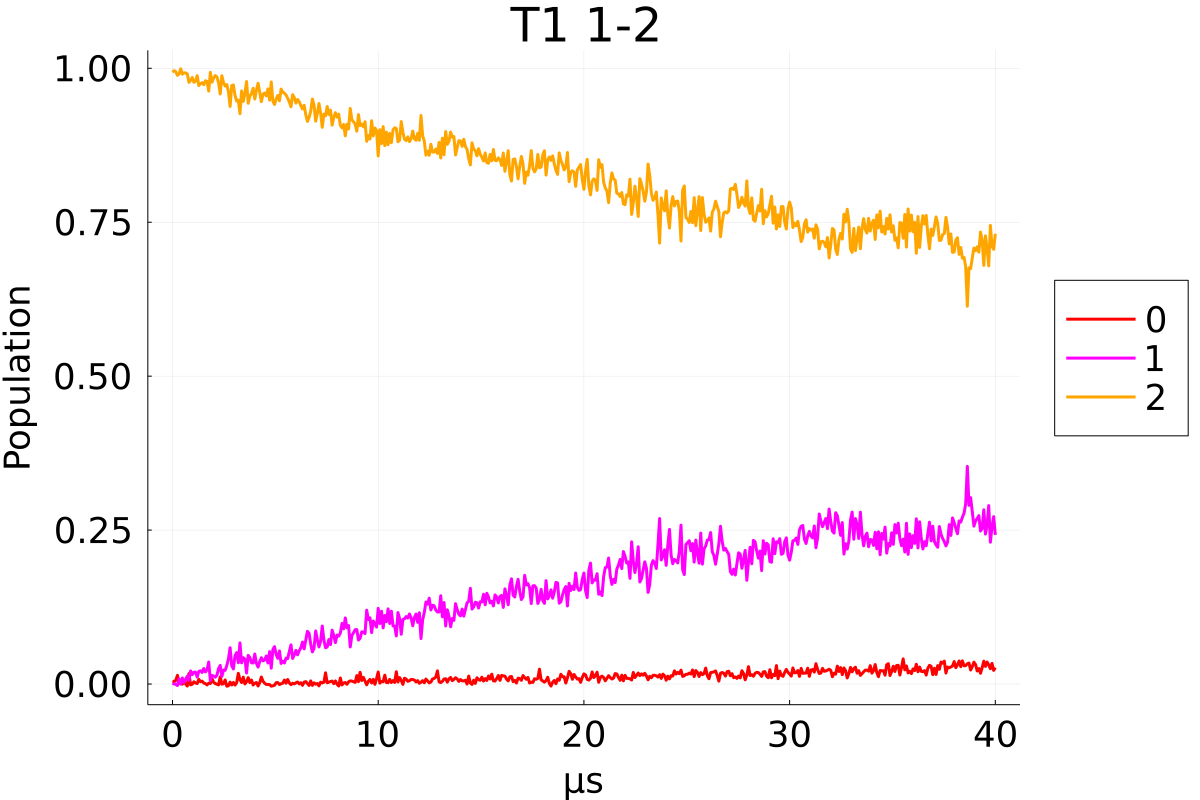}};
  \node(T1-12)[below=of ImgT1,yshift=+1.2cm]{\scriptsize$\pi_{0,1}\rightarrow\pi_{1,2}\xrightarrow[\textrm{free evolution}]{t_{\textrm{dark}}}$};
  \end{tikzpicture}
    \begin{tikzpicture}
    \node (ImgRamsey){\includegraphics[width=0.4\textwidth]{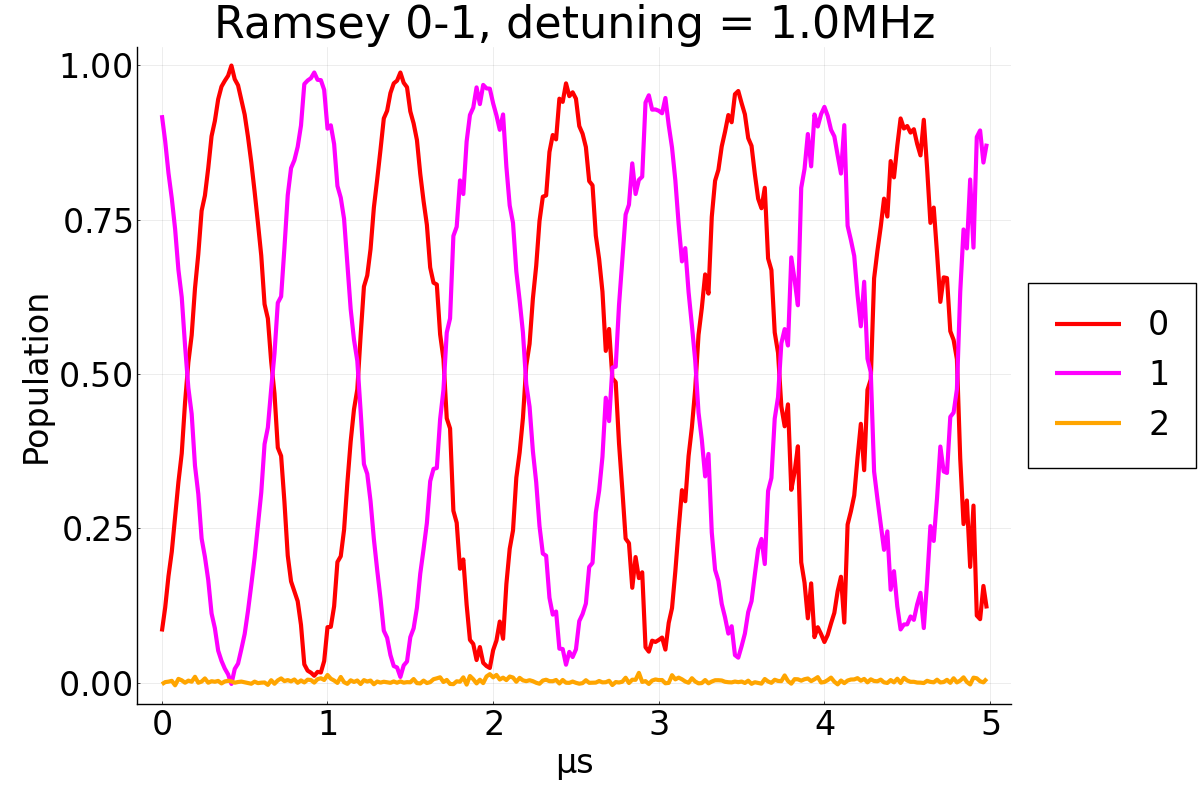}};
    \node(Ramsey)[below=of ImgRamsey,yshift=+1.2cm]{\scriptsize$\frac{\pi_{0,1}}{2}\xrightarrow[\textrm{free evolution}]{t_{\textrm{dark}}}\frac{\pi_{0,1}}{2}$};
    \node(Ramsey)[left=of Ramsey,yshift=-0.05cm,xshift=1.0cm]{\scriptsize Protocol:};
  \end{tikzpicture}
   \begin{tikzpicture}
    \node (ImgRamsey12){\includegraphics[width=0.4\textwidth]{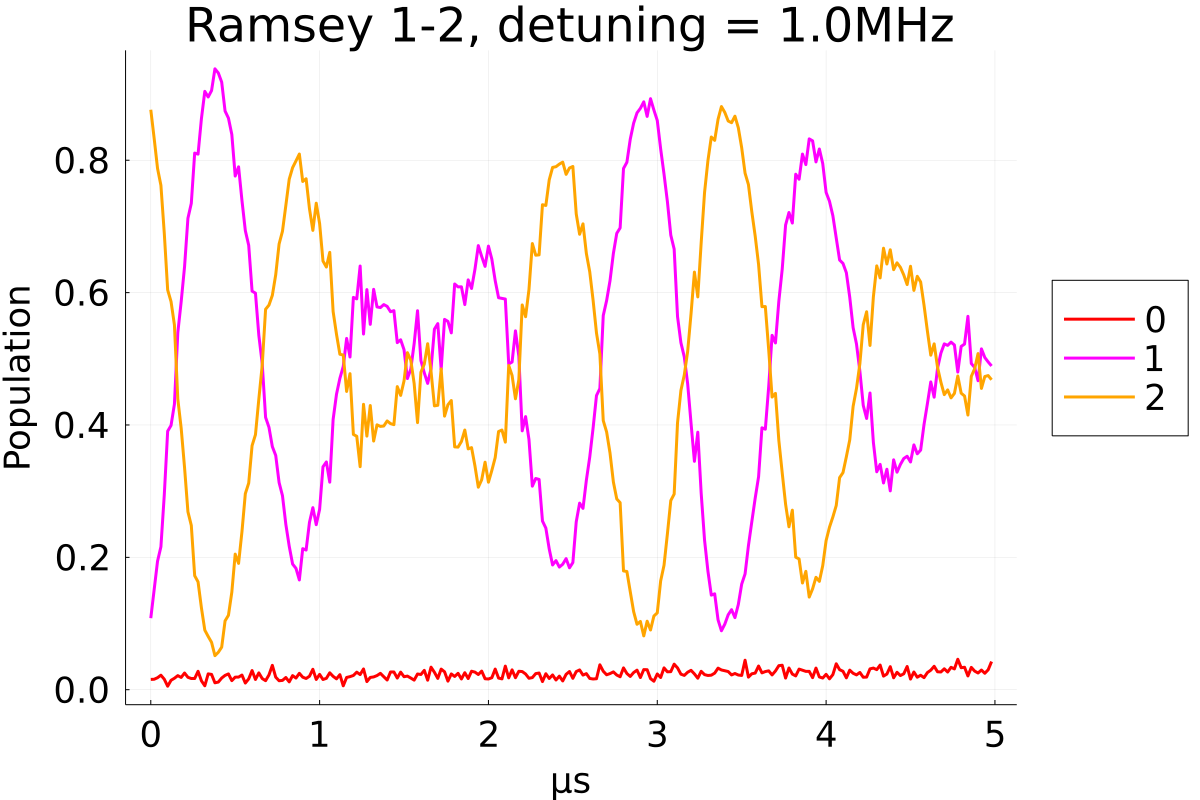}};
    \node(Ramsey12)[below=of ImgRamsey12,yshift=+1.2cm]{\scriptsize$\pi_{0,1}\rightarrow\frac{\pi_{1,2}}{2}\xrightarrow[\textrm{free evolution}]{t_{\textrm{dark}}}\frac{\pi_{1,2}}{2}$};
  \end{tikzpicture}
    \caption{Populations as function of dark time for the energy decay (top) and Ramsey (bottom) experiments. In the protocols, $\pi_{k,k+1}$ and $\frac{\pi_{k,k+1}}{2}$ denote $\pi$ and $\frac{\pi}{2}$ pulses for the $k\leftrightarrow k+1$ transition.}\label{fig:exps}
 \end{center}
\end{figure}

The energy decay experiment is used to estimate the $T_{1,k}$ time scale for state $|k\rangle$ to decay into state $|k-1\rangle$. Here, the qudit is first prepared in state $|k\rgl$ with a series of $\pi$-pulses. The device is then evolved freely (without control pulses) for a given dark time, followed by readout. The dark time is incremented in steps of \SI{80}{\nano\second}, and $1,000$ shots are used for each dark time (i.e., the experiment is repeated $1,000$ times). Populations as functions of the dark time are shown in Figure~\ref{fig:exps} (top panels). We refer to Appendix~\ref{sec:measure-data} for details on how the populations are measured. 

\begin{figure}[thb]
  \begin{center} 
  \includegraphics[width=0.45\textwidth]{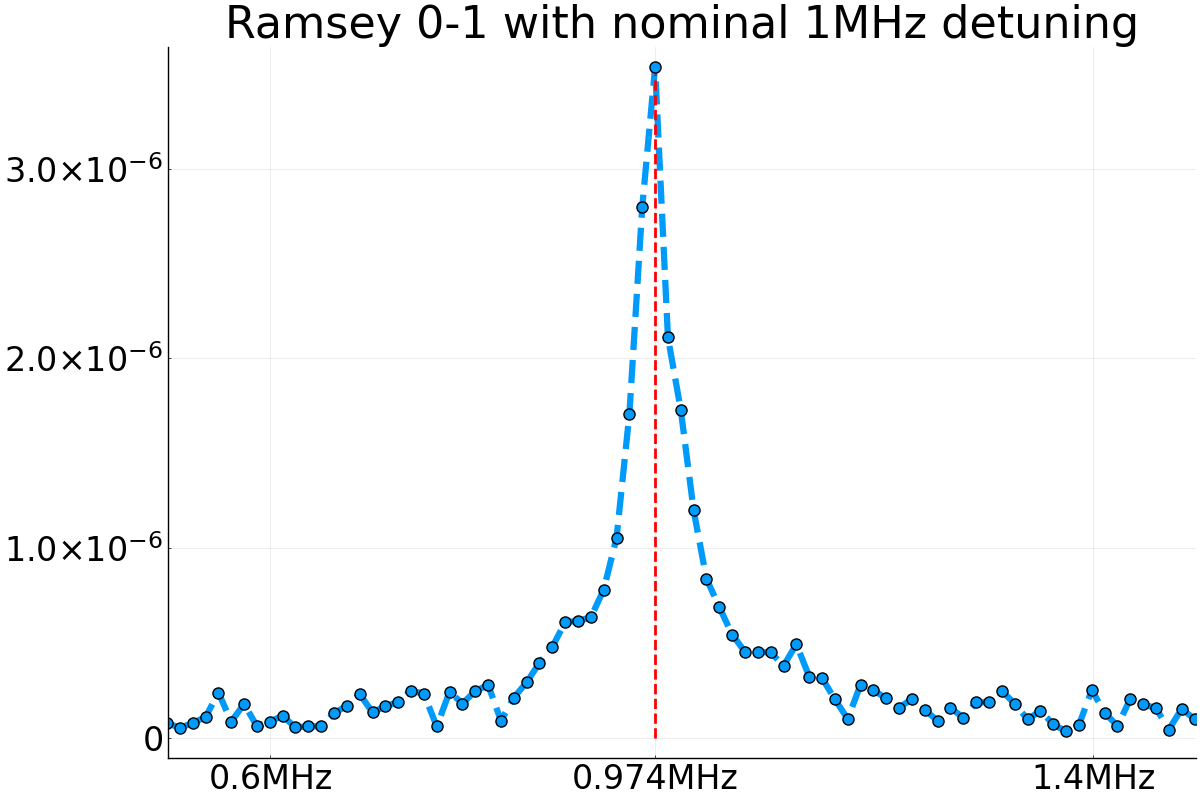}
  \includegraphics[width=0.45\textwidth]{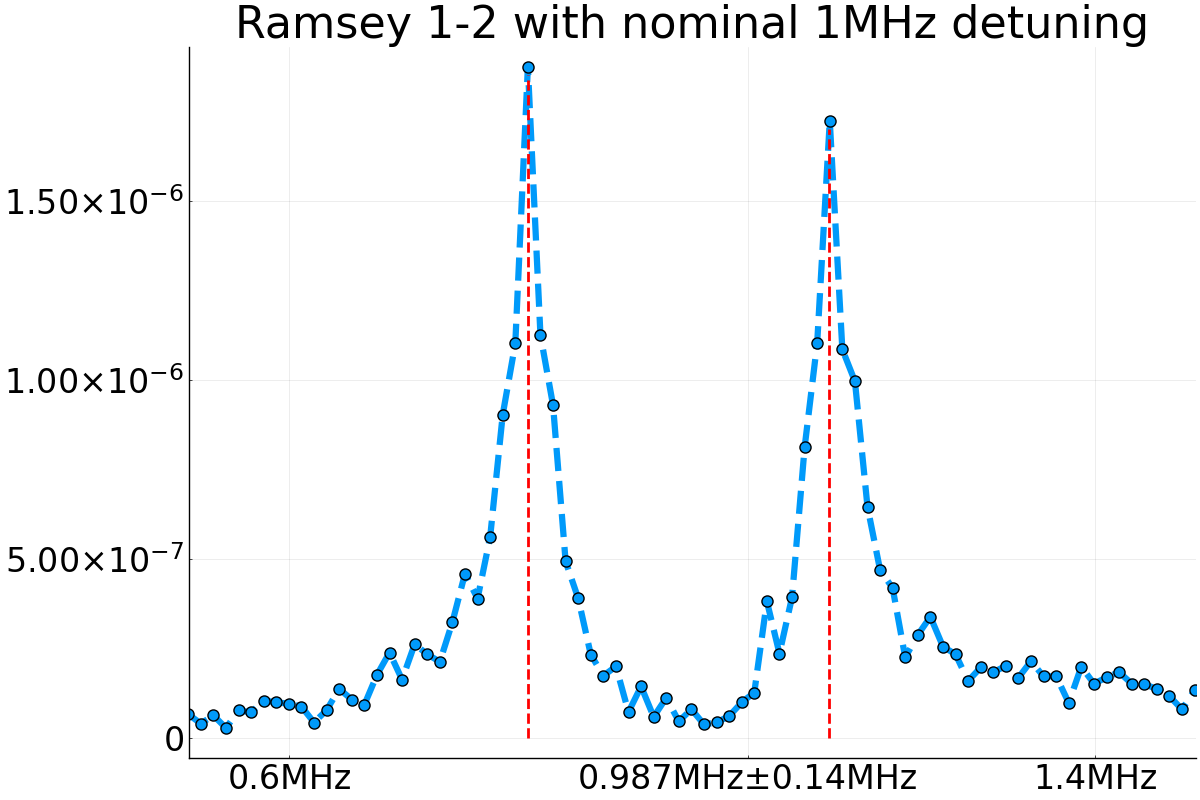}
  \caption{Fourier transform of the Ramsey fringes for the $0\leftrightarrow 1$ and $1\leftrightarrow 2$ experiments. Left: Fourier magnitude of the population of state $|1\rgl$. Right: Fourier magnitude of the population of state $|2\rgl$.\label{fig:ramsey_fft}}
 \end{center}
\end{figure}

The Ramsey $k \leftrightarrow k+1$ experiment is designed to determine the transition frequency $\omega_{k,k+1}$ and the $T^{*}_{2,k}$ dephasing time scale, also known as the combined decoherence time scale~\cite{tempel2011relaxation}. Here, the frequency $\omega_{k,k+1}$ corresponds to transition between the states $|k\rgl$ and $|k+1\rangle$. During each shot, we first apply a series of $\pi$-pulses to prepare the transmon in state $|k\rgl$, followed by a detuned $\frac{\pi}{2}$-pulse to bring the device into a superposition of states $|k\rgl$ and $|k+1\rgl$. The system is then evolved freely (without applying any control pulses) during the dark time $t_{\textrm{dark}}$, after which a second detuned $\frac{\pi}{2}$-pulse is applied. Finally, the resulting state of the system is measured. Each Ramsey experiment is performed for several dark times. Here we discretize the dark time on a uniform grid with a step size of \SI{20}{\nano\second} and perform 1,000 shots for each dark time. To setup the Ramsey experiment, the drive frequency (which determines the amount of detuning) is chosen as the estimated transition frequency from a standard calibration procedure, reduced by a \SI{1}{\mega\hertz} nominal detuning. This procedure resulted in the drive frequencies $\omega_d/2\pi = 3.4476698$ \SI{}{\giga\hertz} and $\omega_d/2\pi = 3.2392576$ \SI{}{\giga\hertz}, for the $0 \leftrightarrow 1$ and $1 \leftrightarrow 2$ Ramsey experiments, respectively (see, e.g.,~\cite{peng2023mathematical} and \cite{qiskit_url_calibration} for details).
Results from the Ramsey experiments are presented in Figure \ref{fig:exps} (bottom panels). In the population data for the $1\leftrightarrow 2$ Ramsey experiment, we note a beating and a phase flip around dark times 1.5 \SI{}{\micro\second} and 5.0 \SI{}{\micro\second}, which we attribute to parity events \cite{riste2013millisecond}. 
These random events occur on a time scale of milliseconds and alternate the transition frequencies of the device through sudden changes in the charge parity. The perturbation of the transition frequency is called the charge dispersion. 
The presence of parity events can be detected by analyzing the population data from a Ramsey experiment. Because the population data oscillates with detuning frequency $\Delta_k=\omega_{k,k+1}-\omega_d$, it is expected that the Fourier spectrum of the population data will exhibit a maximum at the detuning frequency. Typical amplitude spectra of the population data from $0\leftrightarrow 1$ and $1\leftrightarrow 2$ Ramsey experiments are displayed in Figure \ref{fig:ramsey_fft}. While a single peak is observed in the $0\leftrightarrow 1$ the spectrum, two distinctive peaks are present in the $1\leftrightarrow 2$ case. Given these results, we conclude that the charge dispersion is only significant in the latter case. Also see~\cite{PRXQuantum.3.030307}, where measurements were taken on the same device.

\section{Mathematical models\label{sec:model}}



To take Markovian interaction with the environment into account we model the dynamics of the qudit using
Lindblad's master equation~\cite{lindblad1976generators,Nielsen-Chuang}, which can be written on the general form:
\begin{align}
    \dot{\rho} =& -i\left(H(t)\rho - \rho H(t)\right)
    +\sum_{j=1}^{N^2 - 1} \left( {\cal L}_{j} \rho {\cal L}_{j}^\dagger -
\frac{1}{2}\left( {\cal L}_{j}^\dagger{\cal L}_{j}\rho + \rho{\cal L}_{j}^\dagger{\cal L}_{j} \right) \right).
\label{eq:lindblad}
\end{align}
Here, $\rho=\rho^\dagger$ is the density matrix, $H=H^\dagger$ is the Hamiltonian and ${\cal L}_j$ denotes a decoherence operator. These operators are all in $\mathbb{C}^{N\times N}$, with $N$ being the size of the Hilbert space. Only the three lowest energy levels can be reliably measured on the qudit, but to reduce artificial effects from truncation of the Hilbert space, we add a guard level and include the four lowest energy levels of the system in our modeling. Hence, we set $N=4$.

The Hamiltonian is of the form
$H(t)=H_s+H_c(t)$, where $H_s$ and $H_c(t)$ denote the system and control Hamiltonian matrices, respectively. The system Hamiltonian is modeled by the diagonal matrix 
\begin{equation}
\label{eq:system_hamiltonian}
H_{s} =\left(
 \begin{array}{cccc}
0 & 0 & 0 & 0 \\
0 & \omega_{0,1} & 0 & 0 \\
0 & 0 & \omega_{0,1} + \omega_{1,2} & 0\\
0 & 0 &  0 & \omega_{0,1} + \omega_{1,2}+\omega_{2,3} \\
\end{array}
\right),
\end{equation}
where $\omega_{k,k+1}$, for $k=0,1,2$, denotes the angular transition frequency between quantum states $|k\rgl$ and $|k+1\rgl$. 

In the laboratory frame of reference, the control Hamiltonian $H_c(t)$ is of the form \cite{gerry2005introductory}:
\begin{equation}
    H_c(t) = f(t)(a+a^\dagger),
\end{equation}
where $a$ and $a^\dagger$ are the lowering and raising operators. The control function, $f(t)$, is real-valued and can be conveniently written in the form
\begin{equation}
f(t) = 2\,\mbox{Re}\{e^{i\omega_d t} d(t)\} = 2I(t)\cos(\omega_d t)+2Q(t)\sin(\omega_dt).
    \label{eq:control-Hamiltonian-IQ}
\end{equation}
Here, $\omega_d$ is the angular drive frequency, $d(t)$ is a slowly varying envelope function with real and imaginary parts: $I(t)=\textrm{Re}(d(t))$ and $Q(t)=-\textrm{Im}(d(t))$. Here $I(t)$ and $Q(t)$ are called the in-phase and the quadrature components of the control function. These components are typically used as input signals to an IQ-mixer that, in turn, generates the control signal $f(t)$ that is sent to the qudit.

In the following, we will consider two decoherence operators in Lindblad's equation: the decay operator
$\mathcal{L}_1$  and the dephasing operator  $\mathcal{L}_2$. They are defined as
\begin{align}
{\cal L}_{1} &= \begin{pmatrix}
0 & \sqrt{\gamma_{1,1}} & 0 & 0 \\
0 & 0 & \sqrt{\gamma_{1,2}} & 0  \\
0 & 0 & 0 & \sqrt{\gamma_{1,3}} \\
0 & 0 & 0 & 0
\end{pmatrix},\quad
{\cal L}_{2} = \begin{pmatrix}
0 & 0 & 0 & 0\\
0 &\sqrt{\gamma_{2,1}}& 0 & 0\\
0 & 0 & \sqrt{\gamma_{2,2}} & 0 \\
0 & 0 & 0 & \sqrt{\gamma_{2,3}}
\end{pmatrix}.
\label{eq:lindblad_terms}
\end{align}

The parameter $\gamma_{1,k}$ is the decay rate for state $|k\rgl$, with corresponding decay time $T_{1,k}=1/\gamma_{1,k}$. The dephasing rates $\gamma_{2,k}$ are related to the pure dephasing times $T_{2,k}$ through the relations, $\gamma_{2,0}=0$, $\sqrt{\gamma_{2,k}}=\sqrt{\gamma_{2,k-1}}+\sqrt{2/T_{2,k}}, \, k=1,2,3$ (see the appendix of~\cite{peng2023mathematical} for details). The pure and combined decoherence times are related by $1/T^{*}_{2,k} = 1/(2\,T_{1,k}) + 1/T_{2,k}$; see~\cite{tempel2011relaxation,peng2023mathematical}.

The equations above are stated in the laboratory frame of reference. However, as the transition frequencies and the drive frequency are high, typically in the \SI{}{\giga\hertz} range, numerical simulation can be computationally expensive due to sampling requirements of the state vector. To slow down the time scale, we apply the rotating wave approximation (RWA); see for example~\cite{petersson2020quantum}. In a frame rotating with angular frequency $\omega_d$, the Hamiltonian becomes
\begin{equation}
    H_\textrm{rot}(t)=H_s- \omega_d \,a^\dagger a+\widetilde{H}_c(t), \label{eq:hamiltonian_rot}
\end{equation}
where,
\begin{equation}
    \widetilde{H}_c(t)= I(t)(a+a^\dagger)-iQ(t)(a-a^\dagger). \label{eq:rwa_control}
\end{equation}
Henceforth, the RWA is assumed to be valid and all computations will be performed in the rotating frame of reference.

To conform with standard notation for solving differential equations, our calculations are based on the equivalent vectorized formulation of Lindblad's master equation, 
\begin{align}
\frac{d}{dt}\textrm{vec}(\rho) &= - i (I\otimes H_{rot} - H_{rot}^T \otimes I) \textrm{vec}(\rho)+\sum_{j=1}^2 \hat{\mathcal{L}}_j\textrm{vec}(\rho), \label{eq:GLvec}
\end{align}
where 
\begin{equation}
\hat{\mathcal{L}}_j = \mathcal{L}_j\otimes\mathcal{L}_j-\frac{1}{2}\left(I\otimes (\mathcal{L}_j^T\mathcal{L}_j)
+(\mathcal{L}_j^T\mathcal{L}_j)\otimes I
\right).
\end{equation}

A key component for characterization of system parameters is an efficient solver of Lindblad's master equation. In the present work, we only use control functions that are piece-wise constant in the rotating frame of reference, and the Hilbert space is low-dimensional. A highly efficient approach for integrating Lindblad's master equation is therefore through matrix exponentiation. As we shall see below, the deterministic characterization of parameters in Lindblad's equation leads to an optimization problem. To compute gradients of the associated objective functions, we use the automatic differentiation (AD) framework provided by the {\tt{Zygote.jl}}~\cite{innes2018don} package. All numerical techniques for system characterization in both deterministic and Bayesian frameworks that are described in this paper are provided in the {\tt{GLOQ.jl}} \cite{GLOQ} open source software.



\paragraph{Parity event modeling.} 
The parity event in the $1\leftrightarrow 2$ transition frequency is modeled by
\begin{align}
    \omega_{1,2} = \bar{\omega}_{1,2} + p\, \epsilon_{1,2}, \quad p\in\{-1,1\}, 
\end{align}
where $\bar{\omega}_{1,2}$ is the average $1\leftrightarrow 2$ transition frequency, $\epsilon_{1,2}$ is the charge dispersion and $p\in\{-1,1\}$ is a discrete random variable, called the parity, taking values $\pm 1$ with equal probability \cite{riste2013millisecond}. We define the frequencies corresponding to the positive and negative parities as
\[
\omega^\pm_{1,2}=\bar{\omega}_{1,2}\pm\epsilon_{1,2},
\]
and note that  $\omega_{1,2}^\pm$ are the frequencies corresponding to the two peaks in the right panel of Figure \ref{fig:ramsey_fft}. Since $p$ has zero mean, the average $1\leftrightarrow 2$ transition frequency and the charge dispersion are \mbox{$\bar{\omega}_{1,2} = \frac{1}{2}(\omega_{1,2}^++\omega_{1,2}^-)$} and $\epsilon_{1,2}=\frac{1}{2}(\omega_{1,2}^+-\omega_{1,2}^-)$, respectively.

Parity events have been reported to occur on a time scale of milliseconds~\cite{riste2013millisecond}. This time scale is much longer than the duration of a single shot of the experiments, which typically only requires a few microseconds. Although a parity event could occur during a single shot, the disparate time scales indicate that this would be unlikely. On the other hand, the state population is measured by averaging over $1,000$ shots. In our experiments the wait time between successive shots for the same dark time varies between \SI{0.05}{\milli\second} and \SI{0.1}{\milli\second}. As a result it is likely that about half of the shots are performed for each parity. To account for both parities we solve Lindblad's master equation twice, once with $\omega_{1,2}^+$ and once with $\omega_{1,2}^-$, resulting in the density matrices $\rho^+$ and $\rho^-$, respectively. The average of these density matrices is then used in the characterization, described below.



\section{Deterministic characterization}\label{sec_least-squares}

In this section, we seek deterministic point estimates of the device parameters ${\bm \theta}$. This is achieved by minimizing an objective function that represents the mismatch between the results of Lindblad simulations and the experimental data. 
The minimization problem may be solved by a deterministic algorithm, such as the L-BFGS algorithm that we consider here, or a stochastic algorithm that uses randomness to find the minima of the mismatch function. We note that such a characterization technique is deterministic in the sense that it does not account for the uncertainty in the measurements and hence delivers point estimates (or point values) of the quantum device parameters.

Let $P_k(t_j)$ denote the experimentally determined population of state $|k\rgl$ at dark time $t_j = j\Delta t$, with $j=1,2,\ldots,N_T$, where $\Delta t>0$ is the step size, and $N_T$ is the number of time increments. 
%
%
Let the simulated population of state $|k\rgl$ at dark time $t_j$ be denoted by 
\[ 
\hat{P}_k(t_j; \bm{\theta}) = \langle k | \rho(t_j; \bm{\theta}) | k \rangle,\quad k=0,1,2.
\] 
Here, the argument $\bm{\theta}$ holds the parameters (transition frequencies and decoherence times) in Lindblad's equation, 
%
%
\[
{\bm \theta}=({\omega}_{0,1},{\bar{\omega}}_{1,2},\epsilon_{1,2},{\gamma}_{1,1},{\gamma}_{1,2},{\gamma}_{2,1},{\gamma}_{2,2}).
\]
We consider the following optimization problem,
\begin{equation}
\min_{\bm\theta} J(\bm\theta),\quad
    J({\bm \theta}) = \sum_{e=1}^4\sum_{n=0}^2 \Delta t_e \sum_{j=0}^{N_{T,e}} \big(\hat{P}_n(t_j;{\bm \theta})-P_n(t_j)\big)^2,
\end{equation}
where $J$ is the mismatch function. The sum over "$e$" refers to the Ramsey and energy decay experiments, for both the $0\leftrightarrow 1$ and the $1\leftrightarrow 2$ transitions, i.e., four experiments in total. For the energy decay experiments, we take $\Delta t_{dec} = 80$ ns and $N_{T,dec} = 500$. For the Ramsey experiments, we take $\Delta t_{ram} = 20$ ns and $N_{T,ram} = 250$.

Box constraints are imposed on the parameters in ${\bm \theta}$ to numerically solve the optimization problem. The box constraint for $\omega_{0,1}$ is $3,448.7 \pm 1$ MHz and for $\bar{\omega}_{1,2}$ it is $3,240.3 \pm 1$ MHz. For the charge dispersion, $\epsilon_{1,2}$, we use $140 \pm 125$ kHz. Corresponding to the $T_1$-decay times, we impose the constraints $3.33 < \gamma_{1,k} < 100\ \mbox{kHz}$. The $T_{2,k}$-decay times are constrained through the coefficients $\gamma_{2,1}=2/T_{2,1}$ and $\sqrt{\gamma_{2,2}} = \sqrt{\gamma_{2,1}} + \sqrt{2/T_{2,2}}$, using the bounds 7.14 kHz $< \gamma_{2,1} < 10$ MHz ($2 < T_{2,1} < 280\ \mu\mbox{s}$) and $50 < \gamma_{2,2} < 500$ kHz.

The constrained optimization problem is solved using the multilevel {\tt{fminbox()}} option in the {\tt{Optim.jl}} package~\cite{mogensen2018optim}, in which a barrier penalty term in added to $J({\bm \theta})$ to impose the box constraints. The penalized problem is then solved by the L-BFGS method. The results of the deterministic characterization are summarized in Table \ref{tab:det}.
\begin{table}[thb]
\begin{center}
\begin{tabular}{|c|c|c|c|c|c|c|c||c|c|}
\hline
\footnotesize{${\omega}_{01}/(2\pi)$ [GHz]} &  \footnotesize{${\omega}^-_{12}/(2\pi)$ [GHz]}  & \footnotesize{${\omega}^+_{12}/(2\pi)$ [GHz]} & \footnotesize{$T_{1,1}$ [$\mu$s]}& \footnotesize{ $T_{1,2}$ [$\mu$s]} & \footnotesize{$T_{2,1}$ [$\mu$s]} & \footnotesize{$T_{2,2}$ [$\mu$s]} \\
\hline
\hline
3.448646 & 3.240105 & 3.240403 & 258.39 & 100.79 & 38.44 & 29.94\\
\hline
\end{tabular}
\caption{System parameters resulting from the deterministic characterization.}\label{tab:det}
\end{center}
\end{table}

\begin{figure}[thb]
\centering
  \includegraphics[width=0.45\textwidth]{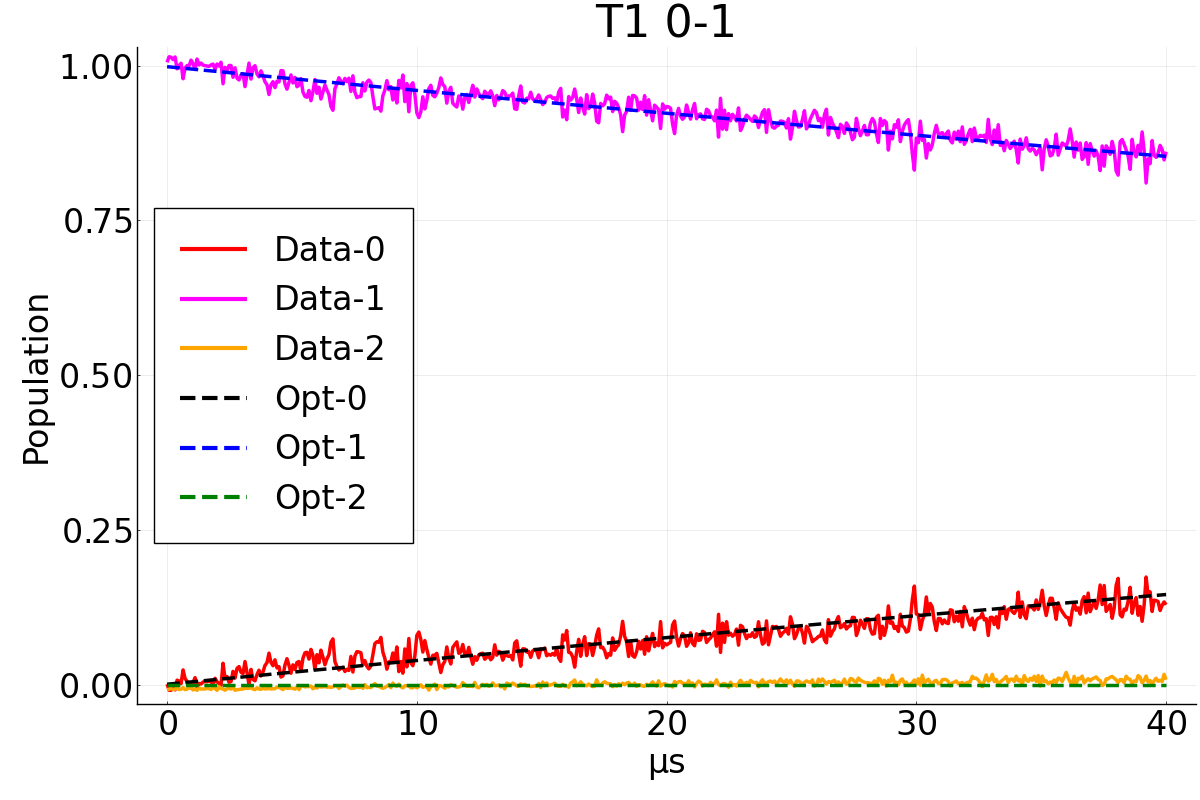}
    \includegraphics[width=0.45\textwidth]{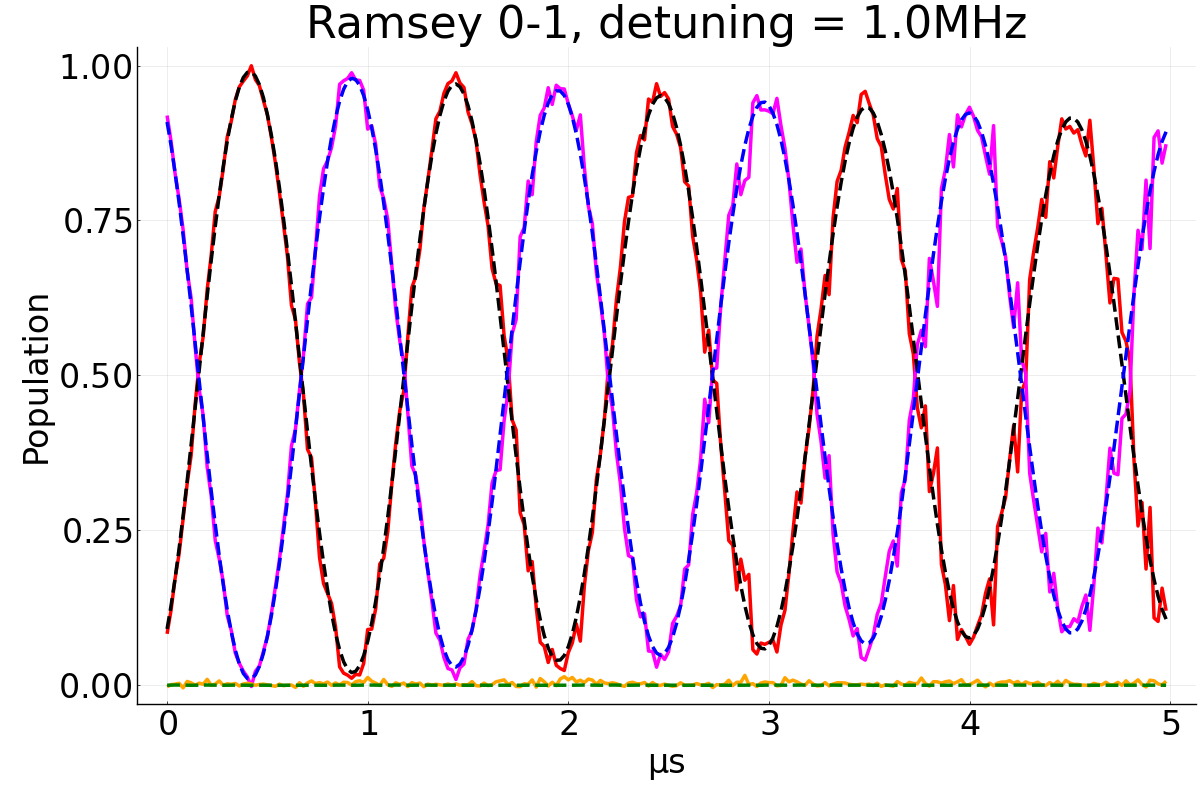}
  \includegraphics[width=0.45\textwidth]{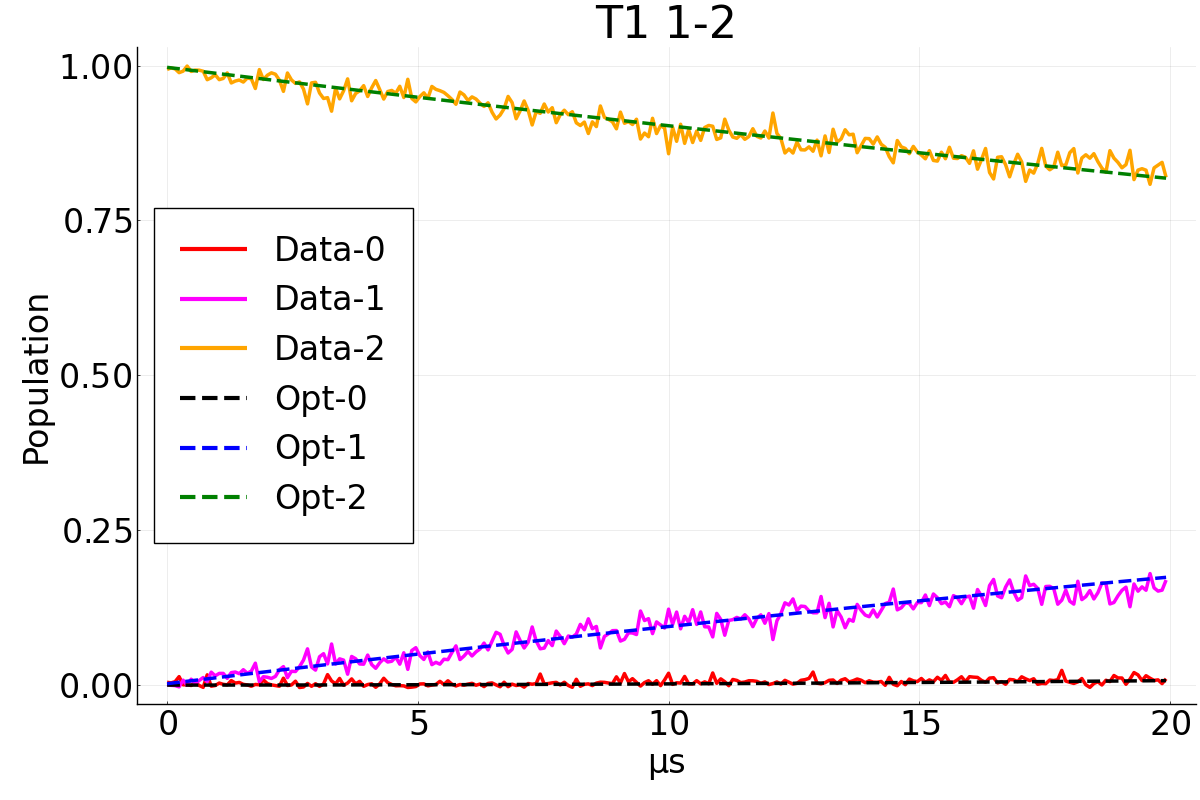}
  \includegraphics[width=0.45\textwidth]{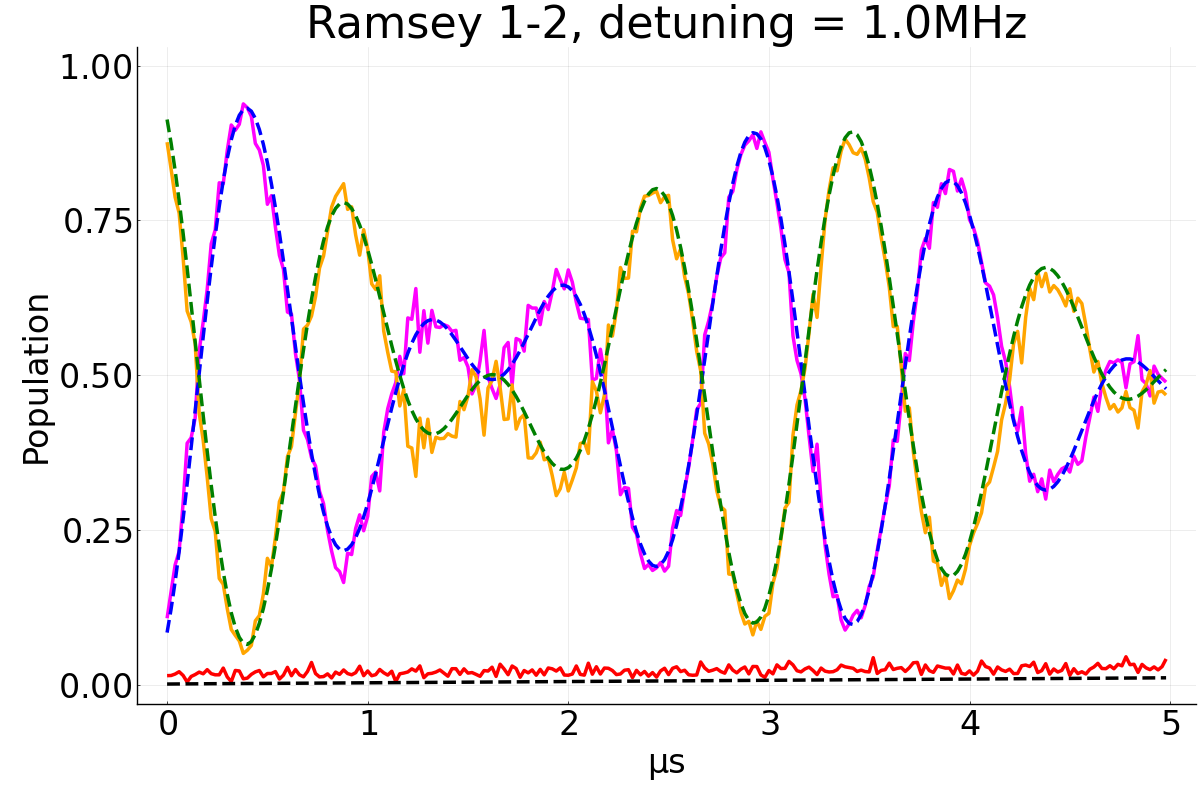}
\caption{Comparison between experimental data (solid lines) and Lindblad simulations (dashed lines), using the parameters from the deterministic characterization.}\label{fig:det_vs_exp}
\end{figure}

Experimental population data and the results from solving Lindblad's equation with the optimized parameters are shown in Figure \ref{fig:det_vs_exp}. We note very good qualitative agreement, including the beatings and phase flips in the Ramsey $1\leftrightarrow 2$ experiment. We emphasize that these features cannot be captured if a single value of $\omega_{1,2}$ is used in the Lindbladian model.

\section{Bayesian characterization}\label{sec:bayesian}

Bayesian characterization is employed to capture the randomness in the device parameters.  The Bayesian approach (see, e.g., \cite{kaipio2006statistical}) starts by assuming that the system parameters ${\bm \theta}$ are stochastic quantities, characterized by a joint probability distribution that is available to us only through a set of randomly corrupted observations. In the present work, we consider the transition frequencies to be random variables with unknown probability distributions and make the simplifying assumption that the decoherence times are known parameters, using the values given in Table \ref{tab:det}. Furthermore, since $T_{1,k}$ is assumed to be known, we only consider data from the Ramsey experiments. 
We note that this simplifying assumption is not restrictive in the sense that the proposed methodology can be applied also in the case where the decoherence times are modeled as random variables. 

Let ${\bm \theta}=(\omega_{0,1},\omega_{1,2}^+,\omega_{1,2}^-)$ be the vector of unknown parameters, and let $D$ denote the population data set collected from the experiments. The goal of the Bayesian characterization is to infer the posterior distribution $\Pr({\bm \theta} | D)$ based on Bayes' rule \cite{Bayes:1763},
\begin{align}
    \Pr({\bm \theta}| D ) = \frac{\Pr( D |{\bm \theta}) \Pr({\bm \theta}) }{\Pr(D)}, \quad \Pr(D) = \int \Pr( D |{\bm \theta}) \Pr({\bm \theta})\, d{\bm \theta}. \label{eq:Bayes}
\end{align}
Bayes' rule relates the posterior distribution of parameters to the product of the prior distribution ${\Pr}({\bm \theta})$, that expresses our belief about $\bm \theta$ before observing the data, and the conditional distribution $\Pr( D |{\bm \theta})$ of the observed data given $\bm \theta$, referred to as the likelihood. The normalization factor ${\Pr}(D)$ in \eqref{eq:Bayes} is referred to as the evidence. 
Given a prior and a likelihood model, a variety of well-established computational tools is available to compute Bayes' posteriors, ranging from Markov Chain Monte Carlo (MCMC) (see, e.g., \cite{MCMC:04}) when the number of parameters is small to variational approximations (see, e.g., \cite{VI:17}) for problems involving many parameters and large data sets.

\subsection{Likelihood function}
A major step in Bayesian inversion is the selection of the likelihood model, based on the structure of the measurement noise. In the present work, we consider a multivariate Gaussian likelihood built under the assumption that the random noise in measurements is additive and normally distributed around zero.

Let $P^{(k)}_n(t)$ be the measured population of state $n \in \{0,1,2\}$, taken at dark time $t$ in the Ramsey $k \leftrightarrow k+1$ experiment, for $k=0,1$. We collect measurements corresponding to many dark times: $t_j=j \Delta t$, for $j=1,2,\ldots,N_T$. Note that each population at a dark time is obtained by averaging over many shots; see Appendix~\ref{sec:measure-data}. 
Further, let $\hat{P}^{(k)}_n(t_j;\bm{\theta})$ be the corresponding simulated populations obtained by the Lindblad model with parameters $\bm{\theta}$. 
We recall that the goal of Bayesian inversion is to recover $\bm \theta$ from a set of (noisy) observations $P^{(k)}_n(t_j)$ of $\hat{P}^{(k)}_n(t_j;\bm{\theta})$, with $n \in \{0,1,2\}$, $k \in \{0,1\}$, and $j \in \{1,2,\ldots,N_T\}$. 
To this end, we first build a Gaussian likelihood model, representing the conditional distribution of the observed data given $\bm \theta$, as follows. Assuming the noise is additive as well as independently and normally distributed for each $j,n,k$, we write
$$
P^{(k)}_{n}(t_j) = \hat{P}^{(k)}_{n}(t_j; \bm{\theta}) + \varepsilon_{n,j}^{(k)}, \qquad n=0,1,2, \qquad j=1,2,\ldots,N_T, \qquad k=0,1, 
$$
where each $\varepsilon_{n,j}^{(k)}$ is a normal random variable with zero mean. 
We further assume that the noise is identically distributed for each $n \in \{0,1,2\}$ and each $j \in \{ 1,2,\ldots,N_T \}$, but since the Ramsey $k \leftrightarrow k+1$ measurements have different noise structures for different $k \in \{0,1\}$ (see Figure \ref{fig:det_vs_exp}), we consider two variances $\sigma_{0}^2$ and $\sigma_{1}^2$ for $k=0$ and $k=1$. In other words, we consider
$$
\varepsilon_{n,j}^{(k)} \sim \mathcal{N}(0,\sigma^2_k), \qquad n=0,1,2, \qquad j=1,2,\ldots,N_T, \qquad k=0,1.
$$
This amounts to the likelihood
$$
\Pr(D|{\bm \theta},\bm\sigma) = (2 \pi)^{- 3 N_T} \, \sigma_0^{- 3 N_T} \,  \sigma_1^{- 3 N_T} \, \prod_{k=0}^1\prod_{n=0}^2
\exp\left(-\frac{1}{2\sigma_{k}^2}\sum_{j=1}^{N_T} \left(P^{(k)}_n(t_j)-\hat{P}^{(k)}_n(t_j;\bm{\theta}) \right)^2\right),
$$
where $\bm\sigma = (\sigma_0,\sigma_1)$. 
Following a common practice in Bayesian inversion with Gaussian likelihoods, we consider the reciprocal of the two noise variances, i.e., the precision parameters $\tau_0:= 1/ \sigma_{0}^2$ and $\tau_1 := 1/ \sigma_{1}^2$, as hyper-parameters; they are unknown and need to be sampled, and yet they are not the main parameters of interest. We hence rewrite the likelihood in terms of the precision hyper-parameters,
\begin{align}\label{eq_total-like}
\Pr(D|{\bm \theta},\bm\tau) \propto \tau_0^{3 N_T /2} \,  \tau_1^{3 N_T/2} \, \prod_{k=0}^1\prod_{n=0}^2
\exp\left(-\frac{\tau_{k}}{2} \, \sum_{j=1}^{N_T} \left(P^{(k)}_n(t_j)-\hat{P}^{(k)}_n(t_j;\bm{\theta}) \right)^2\right),
\end{align}
where $\bm\tau = (\tau_0,\tau_1)$. 

\subsection{Prior distributions}

In general, the choice of prior is problem dependent and is often based on expert opinion. For instance, the prior may be informative with uniform or other types of distributions, or in the absence of any prior information, it may be non-informative, i.e., $\Pr(\boldsymbol\theta) \equiv 1$. Here, we employ two different sets of informative priors: 1) uniform priors for the system parameters $\bm \theta$ informed by the deterministic characterization, and 2) conjugate Gamma priors for the precision hyper-parameters $\bm \tau$, motivated by the fact that Gamma priors are conjugate for Gaussian likelihoods, since they amount to posteriors for $\bm \tau$ from the same Gamma distribution family. 

\bigskip
\noindent
{\bf Uniform priors for $\bm \theta$.} 
The prior for the transition frequencies $\bm \theta= (\omega_{0,1},\omega_{1,2}^+,\omega_{1,2}^-)$ are modeled by uniform distributions and selected based on the deterministic characterization. Specifically, we set 
\begin{equation}\label{uniform_prior}
\Pr(\bm\theta) = \prod_{i=1}^3 \Pr(\theta_i), \qquad \theta_i \sim \text{Uniform}(l_i,u_i),
\end{equation}
with the probability distribution functions,
$$
{\Pr}(\theta_i) = 1 / (u_i - l_i),\quad i=0,1,2.
$$
Here, the support bounds $\{ (l_i,u_i) \}_{i=1}^3$ are determined based on the expectations of the transition frequencies, taken from the deterministic characterization summarized in Table \ref{tab:det}, and with a support length of $2 \times 10^3$ kHz. That is, we take
\begin{alignat*}{3}
 (l_1 - \bar{\omega}_{0,1})/2\pi &=- \SI{1}{\mega\hertz},\quad &
 (u_1 - \bar{\omega}_{0,1})/2\pi &= \SI{1}{\mega\hertz},\\
 (l_2 - \bar{\omega}^+_{1,2})/2\pi &= -\SI{1}{\mega\hertz},\quad&
 (u_2 - \bar{\omega}^+_{1,2})/2\pi &= \SI{1}{\mega\hertz}, \\
 (l_3 - \bar{\omega}^-_{1,2})/2\pi &= -\SI{1}{\mega\hertz},\quad&
 (u_3 - \bar{\omega}^-_{1,2})/2\pi &= \SI{1}{\mega\hertz}.
\end{alignat*}

\bigskip
\noindent
{\bf Gamma priors for $\bm \tau$.} 
The priors for the hyper-parameters $\tau_{k}$, with $k=0,1$, in the likelihood function \eqref{eq_total-like} are modeled by Gamma distributions. Specifically, we set
\begin{equation}\label{prior_tau1}
{\Pr}(\bm \tau) = \prod_{k=0}^1 {\Pr}(\tau_k), \qquad \tau_k \sim \text{Gamma}(\alpha_k,\beta_k),
\end{equation}
with the probability distribution functions,
\begin{equation}\label{prior_tau2}
{\Pr}(\tau_k) \propto \tau_k^{\alpha_k-1} \exp(- \beta_k \, \tau_k),\quad k=0,1.
\end{equation}
Here, $\alpha_k>0$ and $\beta_k>0$ are the shape and rate parameters. As we will see in Section \ref{sec:MHG_algorithm}, the above Gamma priors, being conjugate for Gaussian likelihoods, amount to Gamma posteriors for $\bm \tau$. 
Importantly, prior Gamma distributions are supported on the whole positive real line, independent of the values of the shape and rate parameters, and will hence take all possible values of $\tau_0$ and $\tau_1$ into account. Nevertheless, in the absence of a priori knowledge about the precision hyper-parameters, we select small shape and rate values (e.g. $\alpha_k = \beta_k = 0.01$) in our numerical examples. This choice minimizes the effect of the prior for the precision hyper-parameters, because the prior shape and rate parameters have an additive effect on the posterior shape and rate parameters, as we shall see below.



\subsection{A Markov Chain Monte Carlo sampling algorithm}\label{sec:MHG_algorithm}

We employ a Metropolis-Hastings-within-Gibbs sampling strategy (see, e.g., \cite{Gelman_etal:04}) to sample from parameter posteriors. The algorithm, summarized in Algorithm \ref{ALG_MG}, iteratively generates a sequence of samples, forming a Markov chain, whose distribution approaches the target distribution of parameters in the limit. 
%
The algorithm consists of two interactive parts: a Gibbs sampler and a Metropolis-Hastings (MH) sampler. The Gibbs sampler is used to sample the hyper-parameters $\bm\tau$ with a fixed parameter vector $\bm\theta$, and the MH sampler is used to sample $\bm\theta$ with fixed hyper-parameters $\bm\tau$. 

\bigskip
\noindent
{\bf Gibbs sampler.} 
The Gibbs sampler utilizes the Gaussian likelihood
\eqref{eq_total-like} and the Gamma priors \eqref{prior_tau1}-\eqref{prior_tau2} to sample from Gamma posteriors on its precision hyper-parameters $\bm \tau$, as follows. From Bayes' rule,
$$
\Pr(\bm\tau | \boldsymbol\theta, D) 
\ \propto \ \Pr(D | \bm \theta, \bm \tau) \, \Pr(\bm\tau|\bm\theta)
\ \propto 
\ \Pr(D | \bm \theta, \bm \tau) \,\Pr(\bm \tau),
$$
where $\boldsymbol\theta$ is fixed. 
Assuming the two precision hyper-parameters $\tau_0$ and $\tau_1$ are independent, i.e., with $\Pr(\bm\tau | \boldsymbol\theta, D) = \prod_{k=0}^1 \Pr(\tau_k | \boldsymbol\theta, D)$, we get from \eqref{eq_total-like} and \eqref{prior_tau1}-\eqref{prior_tau2}, 
$$
\Pr(\tau_k | {\bm \theta}, D) \propto \tau_k^{3 N_T / 2} \exp \left( - \frac{\tau_k}{2} \sum_{n,j} \left( P^{(k)}_n(t_j)-\hat{P}^{(k)}_n(t_j;\bm{\theta}) \right)^2 \right) \, \tau_k^{\alpha_k - 1} \, \exp (- \beta_k \tau_k), \qquad k=0,1.
$$
Hence, we obtain the following Gamma posteriors on $\tau_k$, with $k=0,1$,
\begin{equation}\label{Gamma_posteriors}
\tau_k \sim \text{Gamma}(\alpha_k^*,\beta_k^*), \qquad \alpha_k^*=\alpha_k + 3 N_T/2, \quad \beta_k^*=\beta_k+ \frac{1}{2} \sum_{n,j} \left(P^{(k)}_n(t_j)-\hat{P}^{(k)}_n(t_j;\bm{\theta}) \right)^2.
\end{equation}
This will be used to generate new samples of $\tau_k$, with $k=0,1$, given a fixed $\boldsymbol\theta$.

%

\bigskip
\noindent
{\bf Metropolis-Hastings sampler.} 
The Metropolis-Hastings sampler \cite{Metropolis:53,Hastings:70} employs the Gaussian likelihood \eqref{eq_total-like} with fixed  hyper-parameters $\bm\tau$ (obtained by the Gibbs sampler) to generate a sequence of samples from the posterior of $\bm\theta$,
$$
\Pr(\bm\theta | \bm \tau, D) \propto  \Pr(D | \bm \theta, \bm \tau) \, \Pr(\bm\theta).
$$
Given a sample value $\bm\theta^{(m)}$, a new sample
$\bm\theta^{(m+1)}$ is generated as follows. A candidate
sample, say $\tilde{\bm\theta}$, is first generated by a proposal distribution $q(\bm\theta^{(m)},\tilde{\bm\theta})$ from the current sample $\bm\theta^{(m)}$. This candidate sample is then accepted with probability (see e.g. \cite{Chib_Greenberg:95})
\begin{equation}\label{acceptance_prob}
\gamma = \min \Bigg\{ 1, \ 
\frac{\Pr(\tilde{\bm\theta} | \bm \tau, D) \
  q(\bm\theta^{(m)},\tilde{\bm\theta})}{\Pr(\bm\theta^{(m)}
  | \bm \tau, D) \ q(\tilde{\bm\theta}, \bm\theta^{(m)})} 
  \Bigg\}
= 
\min \Bigg\{ 1, \ 
\frac
{\Pr(D | \tilde{\bm\theta}, \bm \tau) \
\Pr(\tilde{\bm\theta}) \ q(\bm\theta^{(m)},\tilde{\bm\theta})}{\Pr(D |\bm\theta^{(m)},\bm \tau) \ 
\Pr(\bm\theta^{(m)}) \ q(\tilde{\bm\theta}, \bm\theta^{(m)})} 
\Bigg\}.
\end{equation}
In the present work, we use a uniform random walk proposal to generate a new sample $\tilde{\bm\theta}$ from a current sample $\bm\theta^{(m)}$:
\begin{equation}\label{proposal_random_walk}
\tilde{\bm\theta} \sim \text{Uniform} (\bm\theta^{(m)} - {\bf r}/2, \,
\bm\theta^{(m)} + {\bf r}/2),
\end{equation}
where ${\bf r} = (r_1, r_2, r_3) \in {\mathbb R}_+^3$ is a support vector, to be selected so that the proposal distribution is neither too wide,
nor too narrow. A too wide proposal would result in an acceptance rate close to zero, and the chain would rarely move to a different
sample. A too narrow proposal would result in an acceptance rate close to one, but the generated samples would not cover the support of posterior. 
With a new sample $\tilde{\bm\theta}$ generated by uniform distributions \eqref{proposal_random_walk} centered around the current sample $\bm\theta^{(m)}$, we will get
$$
q(\bm\theta^{(m)},\tilde{\bm\theta}) =
q(\tilde{\bm\theta},\bm\theta^{(m)}) = \frac{1}{r_1 \, r_2 \, r_3}.
$$
Such a symmetric proposal gives the acceptance ratio $\gamma = \min \{ 1, \Pr(\tilde{\bm\theta} | \bm \tau, D) / \Pr(\bm\theta^{(m)} | \bm \tau, D) \}$. This implies that after generating a new sample $\tilde{\bm\theta}$ by the proposal \eqref{proposal_random_walk}, we will not need to evaluate $q(\bm\theta^{(m)},\tilde{\bm\theta})$ in the numerator, as it cancels out by $q(\tilde{\bm\theta},\bm\theta^{(m)})$ in the denominator. 
This acceptance ratio was used in the original version of the Metropolis
algorithm \cite{Metropolis:53}.

\begin{algorithm}[tbh]
\caption{{\fontsize{11}{12}\selectfont
    Metropolis-Hastings-within-Gibbs Sampling}}
\label{ALG_MG}
\begin{algorithmic} 
\medskip
\STATE {\bf 1.} {\it Initialization}: Select an arbitrary point
$(\bm\theta^{(0)}, \, \bm \tau^{(0)})$, and set $m=0$.

\medskip
\STATE {\bf 2.} {\it Gibbs sampler}: Generate $\tau_k^{(m+1)} \sim \text{Gamma}(\alpha_k^*,\beta_k^*)$, where 
$$
\alpha_k^*= \alpha_k+ 3 N_T/2, 
\qquad
\beta_k^*=\beta_k + \frac{1}{2} \sum_{n,j} \left(P^{(k)}_n(t_j)-\hat{P}^{(k)}_n(t_j;\bm{\theta}^{(m)}) \right)^2.
$$

\medskip
\STATE {\bf 3.} {\it Metropolis-Hastings sampler}: Generate $\bm\theta^{(m+1)}$ as follows:

\begin{itemize}

\item Sample a candidate $\tilde{\bm\theta}$ from
a proposal distribution $q(\bm\theta^{(m)},\tilde{\bm\theta})$, e.g. as in \eqref{proposal_random_walk}.

\item Compute the acceptance probability $\gamma$ as in \eqref{acceptance_prob}, with the likelihood and the prior given in \eqref{eq_total-like} and
\eqref{uniform_prior}, respectively.

\item Set 
$$
\bm\theta^{(m+1)} = \left\{ \begin{array}{l l}
\tilde{\bm\theta} & \qquad \text{if} \ \  \gamma \ge u \sim \text{Uniform}(0,1), \\
\bm\theta^{(m)} &  \qquad \text{otherwise}.
\end{array} \right.
$$
\end{itemize}
 
\medskip
\STATE {\bf 4.} {\it Iteration}: Increment $m$ by 1 and go to step {\bf 2}. 

\end{algorithmic}
\end{algorithm}

\subsection{Bayesian characterization results}
\begin{figure}[thb]
\centering
  \includegraphics[width=0.99\textwidth]{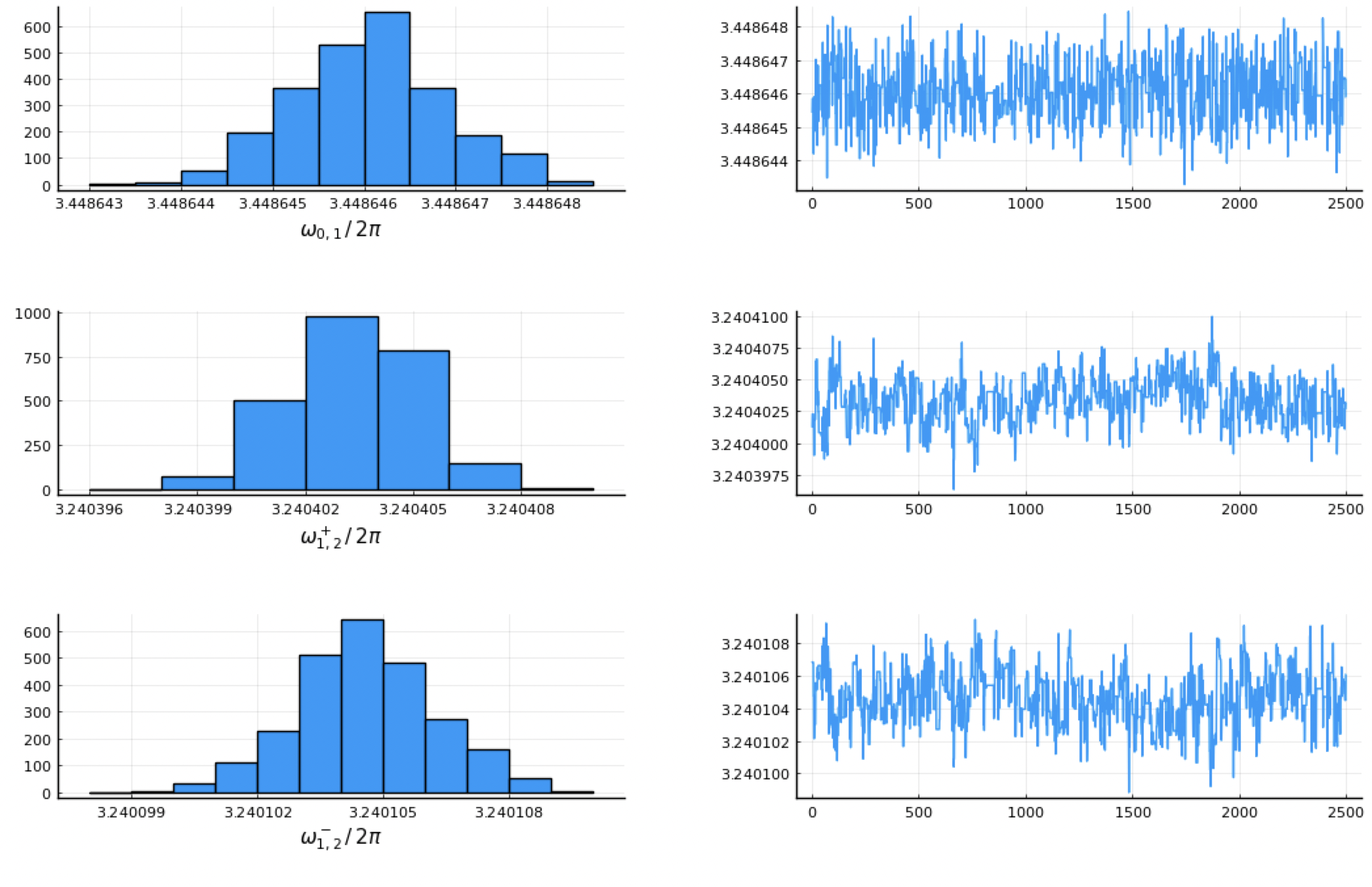}
\caption{Posterior histograms and the trace plots of the Markov chain samples for the transition frequencies drawn by the MH-within-Gibbs sampler.\label{fig:markov-chains_b}}
\end{figure}

We now present the results from the Bayesian characterization using the proposed MH-within-Gibbs sampler (summarized in Algorithm \ref{ALG_MG}). We run the algorithm with $10^4$ iterations and remove the first half of the samples, known as the burn-in period. We also use a thinning period of 2, that is, we discard every other sample in the chain to reduce the correlation between consecutive samples. As a result, we get a total of $2500$ Markov chain samples for the parameters of interest. 
We use the proposal \eqref{proposal_random_walk} with a support $r_1 = r_2 = r_3 = 8$ kHz. 
Figure \ref{fig:markov-chains_b} displays the posterior histograms and the trace plots of the Markov chain samples generated by the algorithm (with an acceptance rate of 27\%). The statistical summaries of the transition frequencies $\omega_{0,1}$ and $\omega_{1,2}^\pm$ are presented in Table \ref{tab:bayesian}. 
It can be noted that the expected values (means) for each of the three parameters are identical to those obtained in the deterministic inversion (displayed in Table \ref{tab:det}). It can also be noted that the variations in each of the three cases, i.e., standard deviations (S.D.) and ranges, are small (on the order of \SI{}{\kilo\hertz}) compared to the expectation of the respective frequency (on the order of \SI{}{\giga\hertz}). 
Moreover, the variations in $\omega_{1,2}^{\pm}$, with a range of 13.6 kHz in $\omega_{1,2}^+$ and 10.3 kHz in $\omega_{1,2}^-$, are higher than the variations in $\omega_{0,1}$ with a range of 5.2 kHz. 
%

\begin{table}[h!]
\begin{center}
\begin{tabular}{|c|c|c|c|c|c|c|c||c|c|}
\hline
&\footnotesize{${\omega}_{01} / 2\pi$} & \footnotesize{${\omega}^-_{12} / 2\pi$}  & \footnotesize{${\omega}^+_{12} / 2\pi$}  \\
\hline
\hline
Mean & $\SI{3.448646}{\giga\hertz}$ & $\SI{3.240105}{\giga\hertz}$ & $\SI{3.240403}{\giga\hertz}$ \\
\hline
S.D. & $\SI{0.8}{\kilo\hertz}$ & $\SI{1.7}{\kilo\hertz}$ & $\SI{1.8}{\kilo\hertz}$ \\
\hline
Range & $\SI{5.2}{\kilo\hertz}$ & $\SI{10.6}{\kilo\hertz}$ & $\SI{13.6}{\kilo\hertz}$ \\
\hline
\end{tabular}
\caption{Statistical summaries, including mean, standard deviation (S.D.), and range of transition frequencies determined by the Bayesian characterization.  \label{tab:bayesian}}
\end{center}
\end{table}

\begin{figure}[thb]
\begin{center}
  \includegraphics[width=0.99\textwidth]{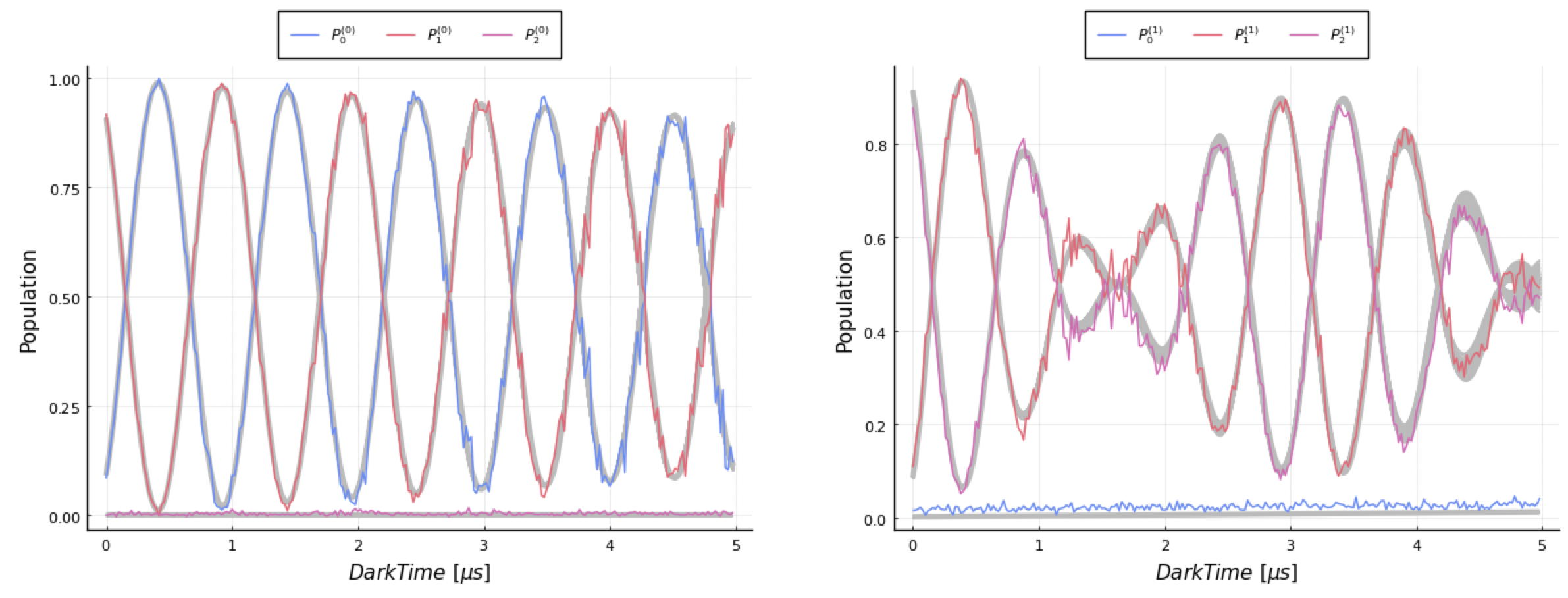}
\caption{Comparison between experimental observations (colored lines) for the Ramsey $0\leftrightarrow 1$ (left) and $1\leftrightarrow 2$ (right) experiments and 200 simulations of Lindblad's master equation (gray lines), where the transition frequencies are sampled from their respective Markov chains. The variable thickness of the gray lines is a result of the variation in transition frequencies.}\label{fig:bayesian-vs-exp_b}
\end{center}
\end{figure}

The MCMC simulations show good qualitative agreement with experimental measurements; see Figure \ref{fig:bayesian-vs-exp_b}, where we overlay 200 simulations of Lindblad's master equation on top of the experimental measurements. Here, the transition frequencies for each simulation are sampled from their respective Markov chains. 

%
%

\section{Conclusions}\label{sec:conclusion}

We have presented a data-driven characterization approach for estimating transition frequencies and decay times in a Lindbladian dynamical model of a superconducting quantum device. The data, collected from the Quantum Device and Integration Test-bed (QuDIT) at Lawrence Livermore National Laboratory, includes parity events in the transition frequency between the first and second excited states. A simple but effective mathematical model, based upon averaging solutions of two Lindbladian models, has been demonstrated to accurately capture the experimental observations. The characterization consists of two stages. First, we perform a deterministic point estimate of all device parameters, followed by a Bayesian characterization approach to capture the randomness in the transition frequencies, driven by the noisy Ramsey measurements. The Bayesian framework utilizes a multivariate Gaussian likelihood with two hyper-parameters, and we have demonstrated that this framework captures the overall structure of the noise in the data. 

Despite a good overall agreement between the experimental measurements and the simulated populations obtained by the inferred parameters, we have observed that some noise variations are not fully recovered by the simulated populations, in particular near the first phase-flip region in the Ramsey $1\leftrightarrow 2$ data; see Figure \ref{fig:bayesian-vs-exp_b}.
We hypothesize two reasons and suggest two potential remedies to be tested. First, improved noise modeling may be achieved by also treating the decoherence times as random variables. Secondly, we note that the noise in both the Ramsey $0\leftrightarrow 1$ and $1\leftrightarrow 2$ cases is not identically distributed and additive, but has a more complex structure. In the $0\leftrightarrow 1$ case, the noise increases as the dark time increases, and in the $1\leftrightarrow 2$ case, the noise increases substantially in the phase-flip regions. This suggests that a Gaussian likelihood model that is built upon the additive noise assumption (as considered here) may not be an optimal model. The construction of other likelihood models, which do not rely on the additive noise assumption, and treating the decoherence times as random variables, will be the subjects of future investigations.

\section*{Acknowledgements}
This work was performed under the auspices of the U.S. Department of Energy by Lawrence Livermore National Laboratory under Contract DE-AC52-07NA27344. This is contribution LLNL-JRNL-850549.
\appendix
\begin{appendices}

\section{Classification of quantum state based on measurements}\label{sec:measure-data}

\begin{figure}[thb]  
\begin{center}
\begin{tikzpicture}
   \node (img){\includegraphics[width=0.75\textwidth,trim={0.0cm 0.28cm 6.0cm 1.0cm},clip]{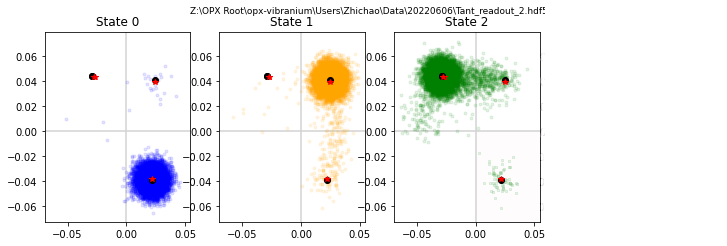}};
   \node[above =of img, node distance=0cm, minimum height=0.05cm,yshift=-1.25cm,xshift=0.25cm] {State $|0\rgl$\hspace{2.25cm}State $|1\rgl$\hspace{2.25cm}State $|2\rgl$};
   \node[below =of img, node distance=0cm, minimum height=0.05cm,yshift=1.0cm,xshift=0.5cm] {$I$\hspace{3.5cm}$I$\hspace{3.5cm}$I$};
   \node[left=of img, node distance=0cm, rotate=90, anchor=center,yshift=-1.2cm] {$Q$};
 \end{tikzpicture}
\end{center}
\caption{A typical measurement result after preparing the device in states $|0\rgl$, $|1\rgl$ and $|2\rgl$, with $\pi$ pulses. $I$: in-phase component; $Q$: quadrature component. \label{fig:IQ-confusion}}
\end{figure}
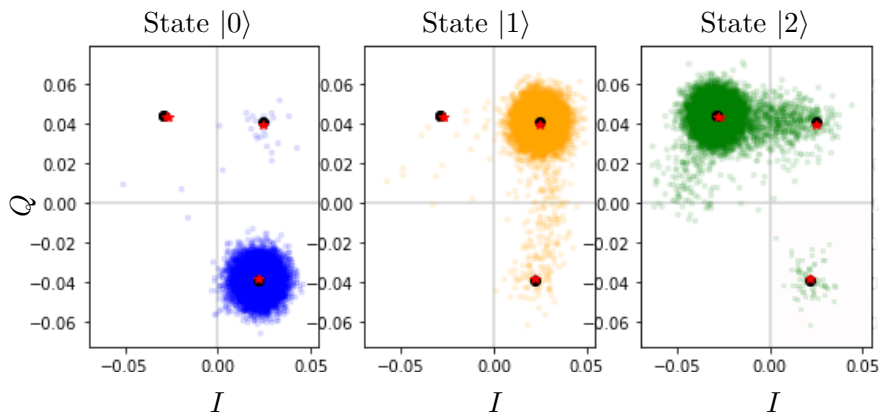

Repeated measurements are used to give statistics of the observables, and it is possible to infer the state of the quantum system from these statistics, for example by using a statistical classifier. Such a classifier can be trained by first preparing the device in different states with a series of $\pi$-pulses, directly followed by measurement \cite{blais2004cavity}. In each shot, the readout pulse is demodulated into an in-phase component, $I$, and the quadrature component, $Q$, as shown in Figure \ref{fig:IQ-confusion}. A typical training set, based on 80,000 shots, is presented in Figure \ref{fig:IQ-confusion}. Each cluster in the $(I, Q)$ plane represents an energy level of the qudit. Using the readouts we train a Gaussian mixture model (GMM) \cite{zhuang1996gaussian,xu2005survey} as a classifier that maps a given $(I,Q)$ coordinate to a vector whose $k$-th element represents the probability of this coordinate belonging to the state $|k\rgl$, i.e., the population $P_k$.

Due to state preparation and measurement (SPAM) errors, the classification of the result is not always accurate. These errors can be mitigated through a confusion matrix, $C$. The confusion matrix can be seen as the transition probability matrix between measured and actual populations. The elements of this matrix, $c_{ij}$, represent the probability of measuring state $|i\rgl$ after the system is prepared in state $|j\rgl$. As an example, the confusion matrix provided by the GMM trained with the data in Figure \ref{fig:IQ-confusion} is
\begin{equation}
C = \begin{pmatrix}
            9.97125\mathrm{e}{-1}& 2.62500\mathrm{e-}{3} &2.50000\mathrm{e}{-4}\\
            1.67500\mathrm{e}{-2}& 9.81250\mathrm{e-}{1} &2.00000\mathrm{e}{-3}\\
            6.12500\mathrm{e}{-3}& 4.33750\mathrm{e-}{2} &9.50500\mathrm{e}{-1}
\end{pmatrix}.
\end{equation}
The populations corresponding to the actual state vector $\psib$ can then be estimated from the measured populations, represented by the state vector $\phib$, by inverting the confusion matrix,
\begin{align}
    \begin{pmatrix}
    |\psi_0|^2\\
    |\psi_1|^2\\
    |\psi_2|^2
    \end{pmatrix}
    = C^{-1}  \begin{pmatrix}
    |\phi_0|^2\\
    |\phi_1|^2\\
    |\phi_2|^2
    \end{pmatrix}.
    \label{eq:error_mitigation}
\end{align}

Each experiment is repeated many times to estimate the average population of the corresponding state. Let the measurement of shot number $s$ result in coordinate $(I_s,Q_s)$. The trained GMM is then used to map this coordinate to a population vector, which is then multiplied by the inverse of the confusion matrix to mitigate measurement errors. Let the resulting population vector be $(P_0^{(s)},P_1^{(s)},P_2^{(s)})^T$. Finally, the {\bf experimentally determined} population of state $|k\rgl$ is taken as the average population over all shots for the same experimental setup,
\[
P_k = \frac{1}{N_s}\sum_{s=1}^{N_s} P_k^{(s)},\quad k=0,1,2.
\]
Here, $N_s$ is the total number of shots. 

\end{appendices}

\bibliographystyle{plain}
\bibliography{ref.bib,ref-70-81.bib}
\end{document}